\pdfoutput=1
\documentclass[reprint,aps,prl,floatfix]{revtex4-2}
\usepackage{graphicx}
\usepackage{dcolumn}
\usepackage{bm}
\usepackage{amsmath}
\usepackage{mathtools}
\usepackage{lineno}
\usepackage[english]{babel}
\usepackage[utf8]{inputenc}
\usepackage{color}
\usepackage{amsmath}
\usepackage{textcomp, gensymb}
\usepackage{upgreek}
\usepackage{physics}
\usepackage[T1]{fontenc}
\usepackage[colorlinks=true, citecolor=magenta, linkcolor=blue, filecolor=magenta, urlcolor=magenta]{hyperref}

\begin{document}

\title{Exploiting epitaxial strained germanium for scaling low noise spin qubits at the micron-scale}

\author{Lucas E. A. Stehouwer$^{1}$}
\author{Cécile X. Yu$^{1}$}
\author{Barnaby van Straaten$^{1}$}
\author{Alberto Tosato$^{1}$}
\author{Valentin John$^{1}$}
\author{Davide Degli Esposti$^{1}$}
\author{Asser Elsayed$^{1}$}
\author{Davide Costa$^{1}$}
\author{Stefan D. Oosterhout$^{2}$}
\author{Nico W. Hendrickx$^{1}$}
\author{Menno Veldhorst$^{1}$}
\author{Francesco Borsoi$^{1 \dag}$}
\author{Giordano Scappucci$^{1 \dag}$}
\email{g.scappucci@tudelft.nl}

\affiliation{$^{1}$QuTech and Kavli Institute of Nanoscience, Delft University of Technology, Lorentzweg 1, 2628 CJ Delft, The Netherlands \\ $^{2}$QuTech and Netherlands Organization for Applied Scientific Research (TNO), Stieltjesweg 1, 2628 CK Delft, The Netherlands \\$^{\dag}$ These authors contributed equally}

\date{\today}

\begin{abstract}
Disorder in the heterogeneous material stack of semiconductor spin qubit systems introduces noise that compromises quantum information processing, posing a challenge to coherently control large-scale quantum devices. 
Here, we exploit low-disorder epitaxial strained quantum wells in Ge/SiGe heterostructures grown on Ge wafers to comprehensively probe the noise properties of complex micron-scale devices comprising of up to ten quantum dots and four rf-charge sensors arranged in a two-dimensional array.
We demonstrate an average charge noise of $\sqrt{S_{0}}=0.3(1)$ $\upmu\mathrm{eV}/\sqrt{\mathrm{Hz}}$ at 1 Hz across different locations on the wafer, providing a benchmark for quantum confined holes. 
We then establish hole-spin qubit control in these heterostructures and extend our investigation from electrical to magnetic noise through spin echo measurements.
Exploiting dynamical decoupling sequences, we quantify the power spectral density components arising from the hyperfine interaction with $^{73}$Ge spinful isotopes and identify coherence modulations associated with the interaction with the $^{29}$Si nuclear spin bath near the Ge quantum well.
We estimate an integrated hyperfine noise amplitude $\sigma_f$ of 180(8)~kHz from $^{73}$Ge and of 47(5)~kHz from $^{29}$Si, 
underscoring the need for full isotopic purification of the qubit host environment.
\end{abstract}

\maketitle

Recent progress with semiconductor spin qubits~\cite{burkard2023semiconductor} has enabled proof-of-principle quantum processors~\cite{xue_quantum_2022,madzik_precision_2022,noiri_fast_2022,mills_two-qubit_2022,wang2024operating} with error rates below the 1\% threshold predicted to enable quantum error correction~\cite{fowler_surface_2012}. However, millions of highly coherent qubits need to be integrated to achieve a realistic quantum advantage~\cite{hoefler_disentangling_2023}. One avenue to improve quantum performance at scale is by advancing material synthesis and fabrication processes to identify and mitigate the dominant noise sources~\cite{de_leon_materials_2021}. In the mature superconducting quantum technology, understanding of noise~\cite{paladino20141} has advanced and shifted from studying single, isolated components to highly integrated and densely connected quantum systems. For example, Google's Sycamore device~\cite{arute_quantum_2019} has served as a test bed for studying correlated noise in a 39-qubit superconducting quantum processor~\cite{harper_learning_2023}, extrapolating the relevant noise models, and exploring the effectiveness of error correction against correlated noise.
In contrast, noise in semiconductor quantum systems has been predominantly studied in isolated components, such as single-charge transistors or individual spin qubits~\cite{yoneda2018quantum,connors2019low,kranz2020exploiting,lodari_low_2021,zwerver2022qubits,paquelet_wuetz_reducing_2023,elsayed2024low,massai_impact_2024}, with recent efforts beginning to explore correlations~\cite{yoneda_noise-correlation_2023} beyond nearest neighbors~\cite{rojas-arias_spatial_2023,DN}. Operating large and highly connected spin qubit systems, in fact, requires a stringent level of electrostatic uniformity. This uniformity is challenged by the disorder introduced by complex semiconductor materials, gate-stacks, and interfaces, which collectively shape the potential landscape of coupled quantum dots.

Recently, Ge/SiGe heterostructures with exceptionally low disorder have been developed by using Ge wafers as substrates for epitaxy, achieving an order-of-magnitude improvement in both dislocation density and two-dimensional hole gas mobility~\cite{stehouwer2023germanium} compared to those grown on Si wafers~\cite{scappucci_germanium_2021} and used for hole spin qubits~\cite{hendrickx2020fast,hendrickx_four-qubit_2021,jirovec_singlet-triplet_2021,hendrickx2024sweet,wang2024operating}.
Here, we exploit such advancements in the semiconductor material stack to comprehensively study and benchmark the noise properties of holes in germanium. We probe simple systems, such as double quantum dots with sensors on the side, as well as more complex spin-qubit devices, integrating ten quantum dots and four sensors in two dimensions. By employing a variety of tools, we assess statistically the noise power spectral density within the same and across different devices on a wafer, measuring under different hole filling conditions. By adopting a single spin as a noise probe at three different qubit sites in a device, we distinguish and quantify the contribution of the three major noise mechanisms in natural germanium qubits: charge noise coupling via spin-orbit interaction and hyperfine interactions with the $^{73}$Ge and $^{29}$Si nuclear spins baths.

\section*{Results}
\subsection*{Charge noise in minimal quantum dot linear arrays}

\begin{figure*}
    \centering
	\includegraphics[width=170mm]{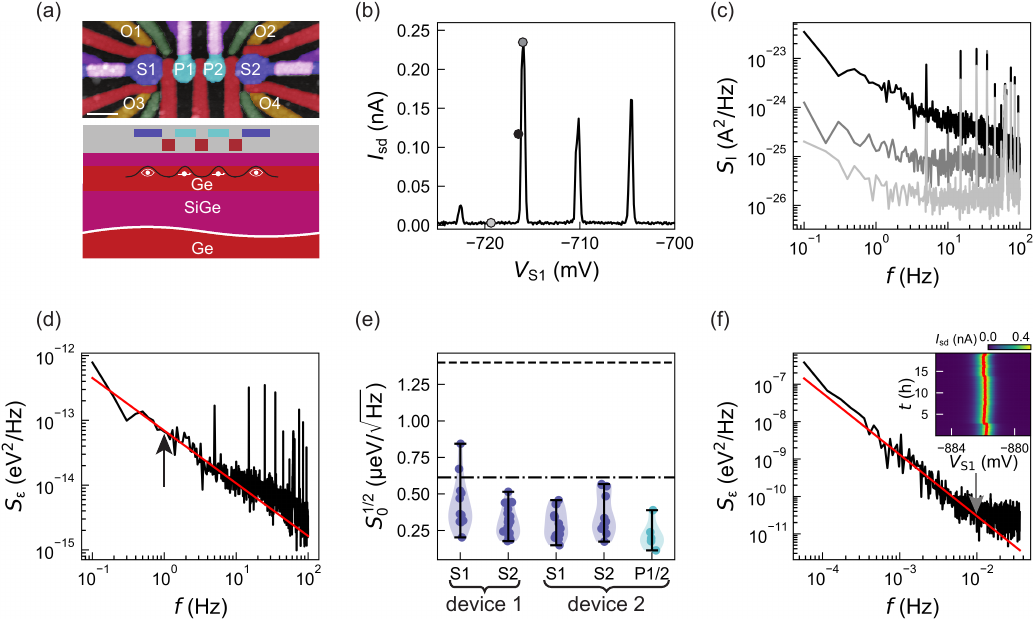}
	\caption{\textbf{Charge noise in minimal quantum dot linear arrays.} (a) False-coloured atomic force microscopy image (top panel) of a device lithographically identical to the measured ones. The device consists of two sensor (dark blue) and two plunger gates (light blue) to define up to four quantum dots, barrier gates (red/green), screening gates (purple), and platinum germanosilicide ohmic contacts (orange). The scale bar is 100 nm. Below we show a side view schematic cutting through the sensor and plunger gates, defining the confining electrostatic potential within the Ge/SiGe heterostructure grown on a Ge wafer. (b) Current $I_\mathrm{sd}$ as a function of sensor voltage $V_\mathrm{S1}$ showing Coulomb peaks from sensor 1 of device 2. Black, grey, and light grey circles mark the flank, top, and blockade region respectively of a Coulomb peak for the measurement of current fluctuations. We use a source drain voltage of $V_\mathrm{sd}=100 \,\upmu\mathrm{V}$. (c) Correspondent current power spectral density $S_{\mathrm{I}}$ from the flank ($\mathrm{V_\mathrm{S1}}=-716.2$ mV), the top ($\mathrm{V_\mathrm{S1}}=-715.9$ mV), and blockade region ($\mathrm{V_\mathrm{S1}}=-717.9$ mV) of a Coulomb peak. (d) Representative charge noise power spectral density $S_{\mathrm{\epsilon}}$ (device 2, S1) from the flank of the Coulomb peak at $\mathrm{V_\mathrm{S1}}=-716.2$ mV. The spectrum is fitted to $S_0/f^{\alpha}$ (red line) using a fit range between 0.1 and 10 Hz to extract the charge noise $\sqrt{S_0}$ at 1 Hz (black arrow). For this spectrum we find a charge noise $\sqrt{S_0}$ at 1~Hz of $0.23(1)$ $\mathrm{\upmu eV}/\sqrt{\mathrm{Hz}}$. (e) Distributions of $S_0^{1/2}$ obtained by repeating charge noise spectrum measurements as in (d) for different hole occupancies. The distribution of acquired charge noise values are represented by violin plots (shaded regions). Whiskers defined by 1.5 times the interquartile range are shown. Measurements are reported for the sensor and plunger gates from two separate devices fabricated on the same low-disorder Ge/SiGe heterostructure on a Ge wafer. As a comparison, the dashed (Ref.~\cite{hendrickx2018gate}) and dash-dotted (Ref.~\cite{lodari_low_2021}) lines show charge noise measurements from similar devices from higher disorder heterostructures, grown on a Si wafer and with the quantum well positioned 22 and 55~nm below the dielectric interface, respectively. (f) Charge noise spectrum (device 2, sensor S1) down to a low frequency of $f=50$ $\upmu$Hz extracted from $\approx$ 18 hour Coulomb peak tracking (inset). The red line in the inset tracks the voltage $V_{\mathrm{S1}}$ for which the current $I_\mathrm{sd}$ of the Coulomb peak reaches its maximum over time. We extract a charge noise of $5.5(9)$ $\upmu$eV/$\sqrt{\mathrm{Hz}}$ at a frequency of 10 mHz (grey arrow) by fitting the data to $S_0/f^{\alpha}$ using a fitting range between 50 $\upmu$Hz and 10 mHz (red line). The uncertainty is 1$\sigma$ from the fit.
    }
\label{fig:one}
\end{figure*}

We begin by characterizing charge noise properties in small quantum dot linear arrays using the flank~\cite{kranz2020exploiting, elsayed2024low, connors2022charge, spence2023probing, paquelet_wuetz_reducing_2023} and the Coulomb peak tracking (CPT)~\cite{kranz2020exploiting, massai_impact_2024} methods, inferring the noise spectrum from about 50 $\upmu\mathrm{Hz}$ to 100~Hz. 
Here, we focus on two nominally identical devices (device 1 and 2) fabricated on the same Ge/SiGe heterostructure on a Ge wafer detailed in Ref.\,\cite{stehouwer2023germanium}, which supports a two-dimensional hole gas with a high maximum mobility of $3.4(1)\times10^ 6$ cm$^2$/Vs and a low percolation density of $1.22(3)\times10^{10}$ cm$^{-2}$. The threading dislocation density is $6(1) \times 10^5$ cm$^{-2}$, nearly an order of magnitude lower than for growth of Ge quantum wells with similar strain~\cite{stehouwer2023germanium} starting from a Si wafer. This improvement is due to the reduced lattice mismatch between the substrate and the epitaxial strained Ge quantum well, which is more than four times smaller for Ge wafers compared to Si wafers~\cite{dismukes_lattice_1964}.

As shown in Fig.\,\ref{fig:one}a, the devices comprise two inner quantum dots (under P1, P2) and two charge sensors at the edges (under S1 and S2), spanning a total distance of about 430~nm. 
Figure\,\ref{fig:one}b shows a representative Coulomb peak series measured in transport on sensor S1 of device 2. Figure\,\ref{fig:one}c illustrates the power spectral density $S_{\mathrm{I}}$ of the current fluctuations as a function of frequency $f$, probed at the top (grey), at the flank (black), and in the blockade region (light grey) of a Coulomb peak (Methods). 
Measurements performed at the flank yield the larger $S_{\mathrm{I}}$, indicating that the noise floor of our setup, probed with the relevant impedance of the load, is sufficiently low to measure the charge noise from the device~\cite{lodari_low_2021}.

We convert the current power spectral density measured at the flank into an energy scale using the slope of the Coulomb peak and its lever arm (here $\approx$ 0.18 eV/V, see Supplementary Fig.\,1).
Figure\,\ref{fig:one}d shows a representative charge noise power spectral density and the associated best fit to the function $S_0/f^{\alpha}$ (red line), with $\sqrt{S_\mathrm{0}}$ the charge noise amplitude at 1 Hz (black arrow in Fig.\,\ref{fig:one}d). The approximate $1/f$ trend of the noise spectrum points towards an ensemble of two-level fluctuators (TLFs) with a wide range of activation energies~\cite{paladino20141, elsayed2024low}. However, we note that, under specific voltage configurations, we observe spectra that deviate from a simple $1/f$ trend (see Supplementary Figs.\,2-5), which could suggest the strong coupling to a single or a few dominating TLFs~\cite{paladino20141, elsayed2024low}. 

We build up statistics by iterating this protocol for different hole occupancies, various single-hole transistors, and two different devices.
We estimate an average charge noise value of $\sqrt{S_\mathrm{0}}$ of $0.3(1)$ $ \upmu\mathrm{eV}/\sqrt{\mathrm{Hz}}$ and $\alpha$ of 0.9(2).
Additionally, we also probe the systems under P1 and P2, by forming a quantum dot in the multi-hole regime under one of the plunger gates, obtaining comparable charge noise values (dark vs. light blue dots in Fig.\,\ref{fig:one}e). 
As highlighted in Fig.\,\ref{fig:one}e, the average charge noise value estimated in this work compares favourably to what reported for previous Ge/SiGe heterostructure implementations. Our value is $\approx 2$ times lower compared to Ge quantum wells buried at the same depth of about 55 nm ($\sqrt{S_\mathrm{0}} = 0.6$ $ \upmu\mathrm{eV}/\sqrt{\mathrm{Hz}}$~\cite{lodari_low_2021}) grown on a silicon substrate and a factor of $\approx 5$ lower than what measured for shallow Ge quantum wells ($\sqrt{S_\mathrm{0}} = 1.4$ $ \upmu\mathrm{eV}/\sqrt{\mathrm{Hz}}$ positioned 22~nm from the dielectric interface~\cite{hendrickx2018gate}).

To further corroborate this result, we extend the characterisation towards lower frequencies by using the CPT method, where the sensor gate voltage is repeatedly swept across a small voltage range around a Coulomb peak~\cite{kranz2020exploiting}. 
We track the Coulomb peak position in time (inset of Fig.\,\ref{fig:one}f for sensor 1 of device 2 and Methods), and, from the fluctuations in its position, we extract the charge noise power spectral density $S_{\mathrm{\epsilon}}$ (Fig.\,\ref{fig:one}f). We determine the value at $f=10$ mHz (grey arrow) by fitting the data to $S_0/f^{\alpha}$ and find a value of $\sqrt{S_{f=10\,\mathrm{mHz}}}=5.5(9)\,\,\upmu\mathrm{eV}/\mathrm{\sqrt{Hz}}$. When we extrapolate the $1/f^{\alpha}$ trend towards 1 Hz, we find $\alpha = 1.64(5)$ and $\sqrt{S_{0}}=0.26(1)$ $\upmu\mathrm{eV}/\sqrt{\mathrm{Hz}}$. The extracted $\alpha$-value of 1.64(5) from the CPT experiment differs with that one  from the flank method ($\alpha=0.9(3)$), possibly due to the presence of drift noise \cite{elsayed2024low}, which we are able to measure due to the long duration ($\approx$18 hours) of the CPT measurement.
$\sqrt{S_{0}}$  is in good agreement with the average charge noise of $0.3(1)$ $\upmu\mathrm{eV}/\sqrt{\mathrm{Hz}}$ that we find using the flank method, confirming our understanding of the system and the reduced noise level in this heterostructure.

\subsection*{Charge noise in a micron-scale 2D quantum dot array}

\begin{figure*}
    \centering
	\includegraphics[width=176mm]{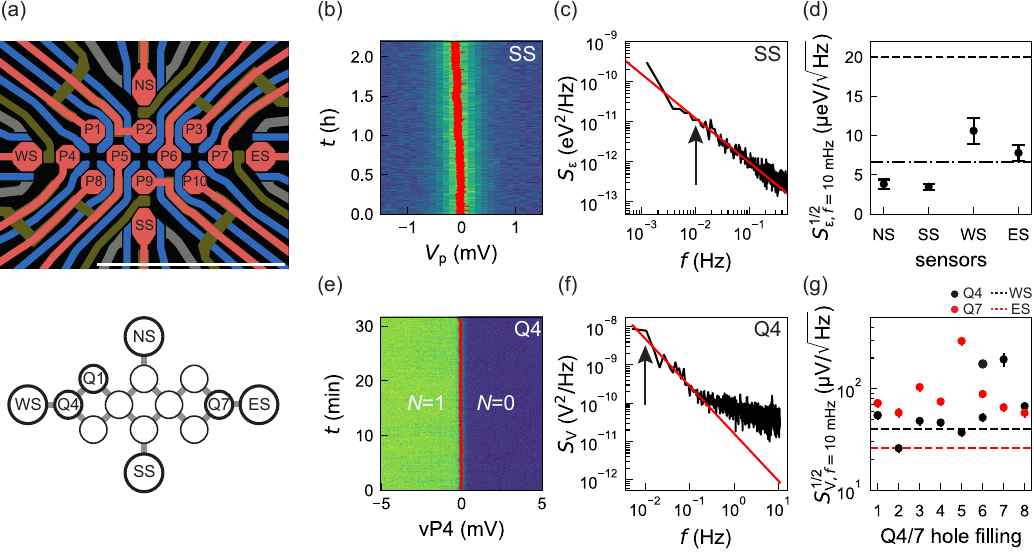}
	\caption{\textbf{Charge noise in a micron-scale 2D quantum dot array.} (a) Schematic of the gate layout of the quantum dot array, hosting 10 quantum dot qubits under plunger gates P1-P10 arranged in a 3-4-3 configuration. Quantum dots are read out by nearby charge sensors NS, ES, SS, WS. Scale bar represents 1 $\upmu$m. The inter-connectivity of the array is shown below. The quantum dots investigated are shown as circles with a thick black line. (b) Two-hour Coulomb peak tracking experiment of the south sensor of the device. (c) Charge noise power spectral density of the south sensor calculated from (b) with $1/f^{\alpha}$ fit between 1 and 100 mHz (red line) and arrow indicating the charge noise at a frequency of 10 mHz. (d) Charge noise from the four sensors and benchmark to the charge noise found from the CPT experiment in Fig.\,\ref{fig:one}f (dash-dotted line) and the charge noise from Ref.\,\cite{massai_impact_2024} (dashed line) at the same frequency (10~mHz). (e) 30-minute repeated loading of the first hole on Q4 by sweeping the virtualised plunger gate (vP4). The red line shows the estimated position for the $N=0$ to $N=1$ charge state transition. (f) Calculated voltage power spectral density from (e) with the extracted noise at 10 mHz (black arrow) from the linear $1/f^{\alpha}$ fit between 10 and 200 mHz (red line). Uncertainties are $1\sigma$ from the fitting procedure. (g) Extracted voltage noise at 10 mHz for the first eigth holes for Q4 (red) and Q7 (black). We do not observe a clear trend of the voltage noise as a function of hole filling. As a comparison we show the voltage noise from the sensors (dashed lines) used to keep track of the charge state of each quantum dot (WS for Q4 and ES for Q7).
    }
\label{fig:two}
\end{figure*}
The improvement in the heterostructure disorder metrics enables exploration and tuning of larger quantum dot architectures.
We probe a two-dimensional (2D) quantum dot array fabricated on the same heterostructure grown on a Ge wafer. We focus on the device shown in Fig.\,\ref{fig:two}a, comprising ten quantum dots arranged in a 3-4-3 configuration and with four rf-charge sensors at the periphery. A similar device design fabricated onto a Si substrate has been recently exploited for studying hopping spin qubit gates in a sparse occupancy~\cite{wang2024operating}. However, here we operate the array in the dense regime, with each quantum dot hosting either one, three, or five holes. 
The outermost charge sensors are 1.5 $\upmu$m apart, a distance that is comparable to the length scale of strain and compositional fluctuations of the heterostructure~\cite{stehouwer2023germanium}. 
This spacing is therefore suitable for investigating the uniformity of noise on a large scale. 

We begin our investigation by performing CPT experiments on the four rf-charge sensors.
Figure\,\ref{fig:two}b shows an exemplary measurement performed on the south sensor SS (see Supplementary Fig.\,6 for the other sensors), which allows us to probe charge fluctuations over a frequency range from 1 mHz to 100 mHz. The corresponding charge noise power spectral density in Fig.\,\ref{fig:two}c is calculated from the Coulomb peak position fluctuations over time, in line with the transport measurements in Fig.\,\ref{fig:one}f (Methods).
By fitting the data with $S_0/f^{\alpha}$, we determine the charge noise values $\sqrt{S_{f=10\,\mathrm{mHz}}}$ at a frequency of $f=10$ mHz (see black arrows in Fig.\,\ref{fig:two}c) and report them in Fig.\,\ref{fig:two}d for the four sensors. The averaged noise across the sensors in this larger array is $6\,\,\upmu\mathrm{eV}/\mathrm{\sqrt{Hz}}$, comparable to the value of $5.5(9)\,\,\upmu\mathrm{eV}/\mathrm{\sqrt{Hz}}$ measured in the smaller linear arrays (dash-dotted line in Fig.\,\ref{fig:two}d) and an exponent $\alpha = 1.21(5)$. These findings suggest that the heterostructure maintains low noise levels both within the same device and across different devices. 
Moreover, a comparison with the best value ($\approx 20\,\upmu\mathrm{eV}/\mathrm{\sqrt{Hz}}$) obtained with CPT measurements on quantum dots defined on a Ge/SiGe heterostructure grown on a silicon substrate~\cite{massai_impact_2024} (dashed line in Fig.\,\ref{fig:two}d) shows a reduction in the noise amplitude by a factor of $\gtrsim 3$.

We also take the step to directly characterise the noise of the inner quantum dots, previously unexplored. We focus on the stability of the outermost quantum dots Q4 and Q7, and probe the susceptibility of noise to local perturbations of the electrostatic environment induced by different charge occupancy and into possible screening effects~\cite{spence2023probing}.

As shown in Fig.\,\ref{fig:two}e for quantum dot Q4, we repeatedly sweep the plunger gate, each time loading the first hole into the quantum dot. We keep track of the transition voltage as shown by the red line in Fig.\,\ref{fig:two}e (see Supplementary Fig.\,7 for further analysis). We quantify the voltage power spectral density $S_\mathrm{V}$ (Fig.\,\ref{fig:two}f) and estimate the voltage noise at a frequency of $f=10$ mHz by fitting to $S_0/f^{\alpha}$. Because of the complexity in determining with accuracy the lever arm of the plunger gates to the quantum dots in this regime, we maintain the metric of the charge noise in voltage, rather than in energy.  
In Fig.\,\ref{fig:two}g we plot the voltage noise as a function of charge filling up to the eighth hole for quantum dots Q4 (black dots) and Q7 (red dots). We do not observe a clear trend of the voltage noise as a function of hole filling, rather we note that the voltage noise fluctuates largely between hole fillings, with average values at $f=10$ mHz of $60(50)$ $\upmu\mathrm{V}/\sqrt{\mathrm{Hz}}$ and $90(70)$ $\upmu\mathrm{V}/\sqrt{\mathrm{Hz}}$ for quantum dots Q4 and Q7, respectively. On average, we obtain a value $\alpha = 1.0(3)$. As a comparison, we also measure the voltage noise of the corresponding charge sensors that were used to sense the charge transitions on the quantum dots (WS for Q4, and ES for Q7) using CPT. We find that the noise of the sensors (black and red dashed line for Q4 and Q7 respectively) is qualitatively comparable to the average noise of the quantum dots.

\subsection*{Charge and hyperfine noise in hole spin qubits}

\begin{figure*}
    \centering
	\includegraphics[width=176mm]{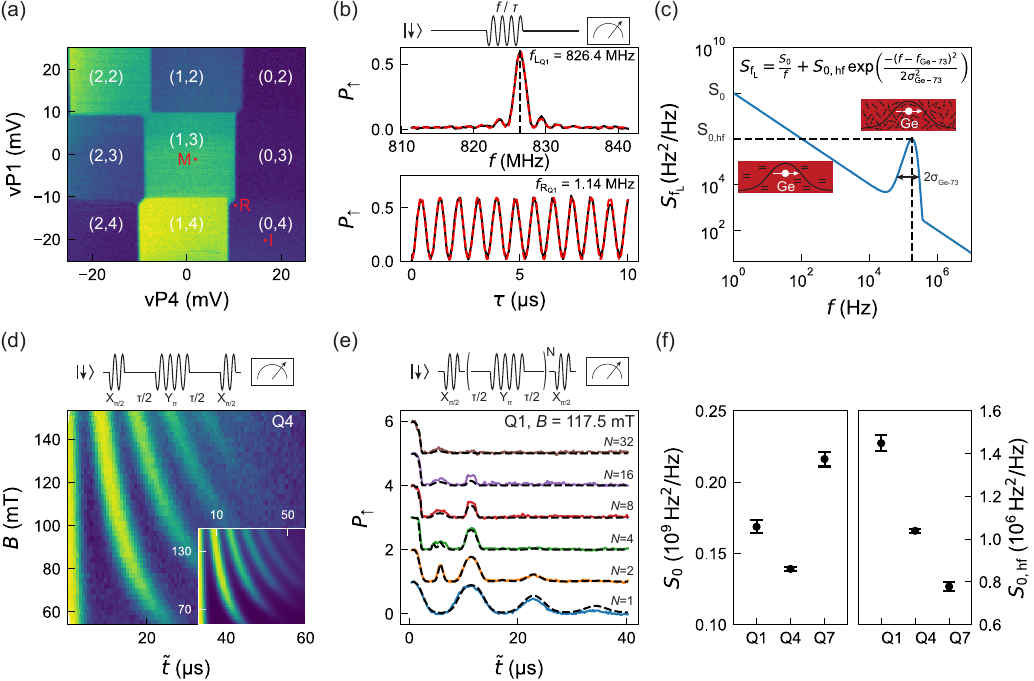}
	\caption{\textbf{Charge and hyperfine noise characterisation using spin echo measurement protocols.} (a) Charge stability diagram for qubit pair Q1-Q4. Labels I, M, and R indicate approximate virtual plunger gate voltages associated with the initialisation, manipulation and read-out stages, respectively. (b) Exemplary Q1 EDSR spectroscopy (top panel) and Rabi oscillations (bottom panel) at $B = 117.5 $ mT. Larmor ($f_\mathrm{L_{Q1}}$) and Rabi frequencies ($f_\mathrm{R_{Q1}}$) are extracted by fitting the data, as discussed in the Methods. Data (best fit) is shown as a black (red dashed) trace. (c) Model of the power spectral density affecting the hole spin qubits. The model consists of a $1/f$ contribution and a Gaussian peak arising from the hyperfine interaction with the $^{73}\mathrm{Ge}$ non-zero nuclear spin. The insets show the coupling of the hole spin qubit to two-level fluctuators (on the left) and to a bath of spinful nuclei (on the right). (d) CPMG-1 experiment for a range of magnetic fields $B$ measured on qubit Q4. Schematic of the pulse scheme on the top panel. The time axis is defined as $\Tilde{t}=\tau+t_\mathrm{\pi}$, where the finite time of the $\mathrm{Y}_{\pi}$ pulse ($t_\mathrm{\pi}=1/(2f_\mathrm{R})$) is taken into account. We normalise the measured signal between 0 and 1 to facilitate the fitting procedure of the model of (c). The inset shows the best-fit to the normalised data using the noise model following Ref.~\cite{hendrickx2024sweet}. Data and fit show excellent agreement. (e) Normalised CPMG-$N$ pulse sequences (schematic on top) for $N=1,\,2,\,4,\,8,\,16$, and 32 $\mathrm{Y}_{\pi}$ pulses for qubit Q1 at $B=117.5$ mT. The black dashed line is the best fit from the noise model in (c) to the data. Each trace is shifted vertically by one unity for clarity. (f) Extracted $S_\mathrm{0}$ and $S_\mathrm{0,hf}$ parameters for qubits Q1, Q4, and Q7 obtained from the data shown in (d), (e) and in Supp. Fig.~8.}
\label{fig:three}
\end{figure*}

We then move from electrical measurements of quantum dot properties to coherent spin experiments to exploit the spin degree of freedom as a sensitive noise probe of the local host environment. 
We begin by demonstrating coherent operations of spin qubits in this Ge/SiGe heterostructure grown on a Ge wafer. We focus our attention on the spin qubit pair confined in the double quantum dot system Q1-Q4. 
Figure\,\ref{fig:three}a illustrates the associated charge stability diagram, obtained by sweeping the two virtual plungers vP4 and vP1~\cite{wang2024operating}. The map reveals well-defined charge regions corresponding to different occupations centered around the ($n_\mathrm{Q4}$, $n_\mathrm{Q1}$) = (1, 3) charge state (with $n_{\mathrm{Q}i}$ indicating the number of carriers confined in quantum dot $i$) regime in which this pair is operated. A magnetic field is applied to Zeeman split the spin states via a one-axis solenoid magnet that is nominally parallel to the sample plane.

We control the qubit pair by pulsing the voltages through the charge stability map. Starting at point I of Fig.\,\ref{fig:three}a, we prepare a two-hole S(0,4) singlet state, then ramp adiabatically to the center of the (1, 3) charge state (point M) preparing a $\ket{\uparrow\downarrow}$ state with an unpaired spin in each quantum dot. Here, we then perform qubit manipulation via electric dipole spin resonance (EDSR), after which we readout the qubit state in R using Pauli spin blockade (PSB). Parallel spin states ($\ket{\downarrow\downarrow}$ and $\ket{\uparrow\uparrow}$) are blocked and mapped into the (1, 3) charge state, while antiparallel spin states ($\ket{\uparrow\downarrow}$ and $\ket{\downarrow\uparrow}$) are transferred into the (0, 4) charge state, resulting in the so-called parity readout (Supplementary Fig. 8)~\cite{Seedhouse2021, Philips2022, Takeda2024_parity}.

We proceed by calibrating the single-qubit gate parameters of each qubit to then exploit Carr-Purcell-Meiboom-Gill (CPMG) sequences to probe the power spectral density of the noise affecting the qubits~\cite{Meiboom1958, Uhrig_alive_2007, Cywinski_enhance_2008, Alvarez_measuring_2011, muhonen_storing_2014, yoneda_quantum-dot_2018}. Figure\,\ref{fig:three}b illustrates EDSR resonant control of Q1 through a microwave burst applied to the plunger P1 at a magnetic field of of 117.5 mT for a varying driving frequency (top) and a varying driving time at the resonant condition (bottom). The Q1 and Q4 Larmor frequencies $f_\mathrm{L_{Q1}} =$ 826.4 MHz and $f_\mathrm{L_{Q4}} =$ 954.5 MHz map to effective \textit{g}-factors of 0.503 and 0.577, respectively.
By taking into account the pronounced anisotropy of the \textit{g}-tensor of planar germanium quantum dots~\cite{jirovec2022dynamics, van2016anisotropy}, we estimate a misalignment angle between the magnetic field direction and the substrate plane of $\approx 3^\circ$ (Methods). Although minimal, the deviation from a perfect in-plane configuration has implications on the sources of noise that affect the qubit.

Our hole spin qubits are, in fact, hosted in a Ge quantum well, which has a natural relative abundance of 7.7 atomic \% of the $^{73}$Ge isotope, which contains a non-zero nuclear spin. The resulting fluctuating Overhauser field couples to the hole spin states via a hyperfine interaction of the Ising type that is highly anisotropic with the magnetic field direction~\cite{Fisher2008, Philippopoulos_2020}.
Following the procedure described in Ref.\,\cite{hendrickx2024sweet}, we quantify the contribution to the qubit decoherence arising from both charge and hyperfine noise by modeling the spectral noise affecting the Larmor frequency ($S_{\mathrm{f_{\mathrm{L}}}}$) with a $S_0/f$ contribution and a Gaussian peak at the precession frequency of the $^{73}$Ge nuclear spins ($f_\mathrm{Ge-73} = \gamma B$, with the expected gyromagnetic ratio $\gamma = 1.48$ MHz/T ), as displayed in Fig.\,\ref{fig:three}c.
We then perform a wide series of CMPG-$N$ experiments, consider the filter function associated with each sequence, and estimate the noise spectral density by varying the number of refocusing $Y_\pi$ pulses $N$, the time in between two pulses and the magnetic field amplitude. 

We begin with a CPMG-1 experiment as a function of the magnetic field on qubit Q4. The resulting map in Fig.\,\ref{fig:three}d manifests a clear collapse-and-revival pattern as a function of time, also known as hole-spin echo envelope modulations~\cite{Wang2012, Philippopoulos2019}, that is dependent on the magnetic field. This phenomenon arises from the sharp noise component at $f_\mathrm{Ge-73} \propto B$ that can be partially filtered only at times of $n/f_\mathrm{Ge-73}$ with $n$ being an integer~\cite{Lawrie2022, hendrickx2024sweet}. Similar patterns have already been observed for GaAs qubits, and the detailed understanding of the qubit interactions with the $^{69}$Ga, $^{71}$Ga and $^{75}$As nuclear spins has enabled protocols that allowed their coherence times to be improved by up to five orders of magnitude~\cite{Bluhm2011, Malinowski2017}.

The inset of Fig.\,\ref{fig:three}d illustrates the fit to the data considering the noise model shown in Fig.\,\ref{fig:three}c, which allows us to estimate the noise component at 1 Hz ($S_\mathrm{0}$), the amplitude of the hyperfine spectral peak ($S_\mathrm{0,hf}$), together with its frequency spread ($\sigma_\mathrm{Ge-73}$). Details of the analysis are discussed in the Methods, while extended measurements and fits as a function of the magnetic field for Q1 are detailed in Supplementary Fig.\,9. 

We extend the noise characterisation using CPMG-$N$ pulse sequences for $N=1,\,2,\,4,\,8,\,16$, and 32 $\mathrm{Y}_{\pi}$ pulses.
The results accompanied by fits at a fixed magnetic field of $B=117.5$ mT ($B=138$ mT) are shown in Fig.\,\ref{fig:three}e (Supplementary Fig.\,9) for qubit Q1.
Using the same approach, we also probe qubit Q7 (see Supplementary Fig.\,9) that is located $\approx 850$ nm from Q4. 

Figure\,\ref{fig:three}f displays the extracted values of $S_\mathrm{0}$ and $S_\mathrm{0,hf}$ for the three qubits in a fixed magnetic field of $B=117.5$ mT, resulting in an average value of $S_\mathrm{0}$ equal to $0.17(3)\times 10^9$ $\mathrm{Hz}^2/\mathrm{Hz}$. 
While nuclear spin effects could in principle also contribute to the low-frequency noise spectrum due to slow nuclear diffusion~\cite{Chekhovich2013, Rojasarias2024}, we find it more plausible to associate the low frequency component $S_\mathrm{0}/f$ to charge noise in the device. The extracted value of $S_\mathrm{0}$, in fact, can be converted to a voltage noise level and compared with our findings on the electrical noise. We assume a spatially homogeneous distribution of uncorrelated fluctuators, and exploit knowledge of the \textit{g}-factor susceptibility to voltage variations in all the surrounding gates~\cite{john_two-dimensional_2024}. Considering traps under the gates the most dominant, uncorrelated noise sources, we can estimate the resulting overall \textit{g}-factor susceptibility of the hole spin qubits as $\frac{\Delta g}{\Delta V} = \sqrt{ \sum_i ({\frac{\delta g}{\delta V_i}})^2} \approx 6.7 \cdot \, 10^{-4} \, \mathrm{mV^{-1}}$, with $\delta g / \delta V_i $ the susceptibility of the g-factor $g$ to a voltage variation on each gate $i$ of the twelve barrier and ten plunger gates controlling the array. 
The associated Larmor qubit fluctuations result then into $\frac{\Delta f_\mathrm{L}}{\Delta V} = \frac{1}{h} \, \frac{\Delta g}{\Delta V} \, \mu_B B$, which for B = 117.5 mT, is $\approx 1.1 \, \mathrm{MHz \, mV^{-1}}$. 
We can then derive an effective voltage power spectral density value at 1 Hz of 
\begin{equation}
 S_\mathrm{V} = \frac{S_\mathrm{0}}{(\frac{\Delta f_\mathrm{L}}{\Delta V})^2} = 140(30) \, \mathrm{\upmu V^2 / Hz,}  
\end{equation}
that results in to an effective voltage noise of $\sqrt{S_\mathrm{V}} = 12(1) \, \mathrm{\upmu V / \sqrt{Hz}}$. This represents a two-fold improvement with respect to what measured for a single qubit in ref.~\cite{hendrickx2024sweet} for a germanium quantum well buried at the similar depth but grown on a silicon substrate.
Moreover, assuming a $\propto 1/f$ framework, the value at 10 mHz ($120(10) \, \mathrm{\upmu V / \sqrt{Hz}}$) lies within the range of what detected from direct quantum dots measurements ($80(60)\, \mathrm{\upmu V / \sqrt{Hz}}$), supporting our hypothesis. 

The hyperfine noise values extracted from the three different qubits are in the same order of magnitude as in Ref.\,\cite{hendrickx2024sweet} for qubits that are not operated on a hyperfine sweet spot (e.g. not in a configuration with the magnetic field pointing in the substrate plane), with qubit to qubit variations that are amplified by the large anisotropy in the sensitivity to hyperfine noise. Exploiting the extracted noise parameters, we estimate an integrated hyperfine noise amplitude of $\sigma_f = 180(8) \, \mathrm{kHz}$, which sets an approximate upper bound for the hyperfine-limited dephasing time of $T_2^* = 1.25(5) \, \mathrm{\upmu \, s}$, qualitatively similar to what measured experimentally in the range of $1 - 2 \, \mathrm{\upmu \, s}$~\cite{john_two-dimensional_2024} (Methods). Performing a similar analysis on the low-frequency component, we predict a charge noise-limited dephasing time of $T_2^* = 3.7(3) \, \mathrm{\upmu s}$ at 117.5 mT, and $44(4) \, \mathrm{\upmu s}$ at 10 mT (Methods).

As, in practice, both noise components act on the qubits at the same time, we further use the model to validate the observed dependence of $T_2^\mathrm{H}$ (envelope decay) with the magnetic field, which, in the investigated magnetic field range, manifests a monotonic increase (shown for Q1 and Q4 in Supplementary Fig.\,10).
Our analysis suggests that for a magnetic field below $\approx 150 $ mT the dominant noise source is the hyperfine interaction with the $^{73}$Ge bath. Coupling of the qubit to charge noise through spin-orbit interaction sets the boundary for $T_2^\mathrm{H}$ at higher magnetic field, with a crossover point that exhibits an optimal $T_2^\mathrm{H}$ of $\approx 40$ $\upmu$s. 
The reduced charge-noise limited coherence times above the crossover is due to the proportionality of the qubit frequency fluctuations with the magnetic field through the g-factor ($\Delta f_\mathrm{L} \propto B$)~\cite{Lawrie2022, hendrickx2024sweet, wang2024operating}.

\subsection*{Hyperfine interaction with \texorpdfstring{\boldmath{$^{29}$}}S{S}i nuclei} 
Because the hole's wavefunction is electrostatically confined near Ge/SiGe interface~\cite{terrazos_theory_2021,wang_modeling_2024}, the interaction with the spinful $^{29}\mathrm{Si}$ nuclei in the barrier may potentially introduce an additional noise source affecting coherence. 

\begin{figure}
    \centering
	\includegraphics[width=\columnwidth]{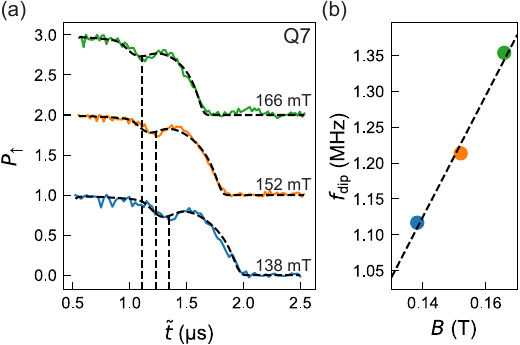}
	\caption{\textbf{Influence of $\mathbf{^{29}}\mathbf{Si}$ nuclear spin on qubit coherence.} (a) Collapse and revival of the qubit Q7 spin state during a CPMG-64 measurement protocol for a magnetic field of 138 (blue), 152 (orange), and 166 (green) mT. The dashed black line shows the best fit when taking into account a second gaussian peak in the noise model from the $^{29}\mathrm{Si}$ non-zero nuclear isotopes. Each trace is shifted upwards by one unit for clarity. A dip in the coherence is observed between 1 and 1.5 $\upmu$s and is attributed to the interaction with the $^{29}\mathrm{Si}$ nuclear spins present in the SiGe barriers. (b) Frequency $f_\mathrm{dip}$ converted from the time of the dip in (a). From the linear fit we extract a gyromagnetic ratio of $8.6(9)$ MHz/T which agrees with the expected 8.465 MHz/T for $^{29}\mathrm{Si}$.}
\label{fig:four}
\end{figure}

As the gyromagnetic ratio of the $^{29}\mathrm{Si}$ nuclei (8.465 MHz/T) is much higher than that of the $^{73}\mathrm{Ge}$ nuclei (1.48 MHz/T), the suppression and revival in the coherence may be visible at a much shorter time scale in a CPMG experiment. 
To isolate this possible decoherence mechanism, we perform a narrow-band noise measurement using a CPMG-64 sequence on qubit Q7 for three magnetic fields, as shown in Fig.\,\ref{fig:four}a. 
In addition to the collapse of the coherence at $ \approx 2\,\upmu\mathrm{s}$ due to the hyperfine interaction with the $^{73}\mathrm{Ge}$ isotopes, we also observe a less pronounced dip between 1 and 1.5 $\upmu\mathrm{s}$. We extract the time associated with this dip ($\Tilde{t}_\mathrm{dip}$) and convert it into a frequency $f_\mathrm{dip}=(2n-1)/2\Tilde{t}_\mathrm{dip}$ (with $n$ indicating the harmonic). We then calculate $f_\mathrm{dip}$ for $n=2$ for the three magnetic fields and use a linear fit to find a gyromagnetic ratio of $8.6(9)$ MHz/T, which agrees with the expected 8.465 MHz/T for $^{29}\mathrm{Si}$ (Fig.\,\ref{fig:four}b). 
We also find qualitative agreement when we expand our noise model with an additional Gaussian peak associated with the $^{29}\mathrm{Si}$ nuclear spins (Methods), confirming that the qubit coherence is also influenced by the interaction with the $^{29}\mathrm{Si}$ nuclear spin bath. From the fit of the three traces, we obtain the average parameters of $S^{\mathrm{Si-29}}_\mathrm{0,hf} = 9(1) \cdot 10^3 \, \mathrm{Hz^2/Hz}$ and $\sigma_\mathrm{Si-29} = 99(20)$ kHz, which results in an integrated hyperfine noise amplitude $\sigma_f$ of 47(5)~kHz and a $^{29}$Si-limited dephasing time of $T_2^* = 4.8(7) \, \mathrm{\upmu s}$ (Methods). 

Given the precision of the noise model to fit the data of Figs.\,\ref{fig:three}d,e and Fig.\,\ref{fig:four}a, we estimate an upper limit on the Hahn decay time $T_2^{\mathrm{H}}$ for the ideal case of perfect isotopic purification of both germanium and silicon. 
To do so, we set the Ge-73 hyperfine noise amplitude $S_{0,\mathrm{hf}}=0$ $\mathrm{Hz}^2/\mathrm{Hz}$ and calculate the qubit coherence considering only the extracted average low-frequency contribution associated to charge noise, $S_\mathrm{0}$ of $0.17(3)\times10^9$ $\mathrm{Hz}^2/\mathrm{Hz}$. We obtain a $T_2^{\mathrm{H}}$ of around 0.4 ms at a small magnetic field of 10 mT, and of 36 $\upmu$s for 117.5 mT, vs. the experimentally detected measurements of 25 $\upmu$s and 36 $\upmu$s for Q1 and Q4 respectively (see Supplementary Fig.\,10). This shows the potential high gain in coherence when isotopically purifying the Ge quantum well as well as the surrounding SiGe barrier layers, and operating the qubits at smaller magnetic fields. 

\section*{Conclusions}
We build a comprehensive understanding of noise in low-disorder Ge/SiGe heterostructures grown on Ge wafers by using a variety of tools applied to increasingly complex devices. 
Canonical characterisation using the flank method in the multi-hole regime shows a low charge noise of 0.3(1) $\upmu\mathrm{eV}/\sqrt{\mathrm{Hz}}$ at 1 Hz, consistently across six quantum dots from two different devices. 
This value is a significant improvement over previous generations of Ge/SiGe heterostructure and sets a benchmark for quantum confined holes in semiconductors ~\cite{hendrickx2018gate,lodari_low_2021,massai_impact_2024}.
Previous studies of strained Ge/SiGe transistors on Si wafers~\cite{simoen_defect-related_2011,simoen_device-based_2019}, as well as Si/SiGe transistors~\cite{hua_threading_2005} and quantum dots~\cite{paquelet_wuetz_reducing_2023}, have linked charge noise to electrically active dislocations originating below the quantum well. Based on these findings, we speculate that charge noise in Ge/SiGe heterostructures grown on Ge wafers may be lower and more uniform than in those grown on Si wafers, due to the significantly reduced density of threading dislocations. Additionally, the reduced topography of the cross-hatch pattern~\cite{stehouwer2023germanium,corley-wiciak_nanoscale_2023} may lead to a more uniform semiconductor-dielectric interface, further improving charge noise uniformity.
Furthermore, as we scale the device in size and complexity, CPT measurements in the multi-hole regime show charge noise that is, on average, uniform across a length scale of 1.5 $\upmu$m within the same device and comparable to the smaller devices. However, when the voltage noise is probed as function of hole occupancy, we observe large fluctuations associated with the specific hole filling. This observation hints to a complex behaviour, where each charge configuration, corresponding to different applied voltages, possibly induces a different coupling to the noise environment surrounding the quantum dot.

We then demonstrate that quantum dots in Ge/SiGe heterostructures on Ge wafers can effectively support qubits. We perform CPMG pulse sequences on three separate qubits and decouple the electrical noise from the magnetic noise by exploiting a theoretical model. We use the extracted parameters to quantify their contribution to decoherence in a range of magnetic fields and to make a prediction of the coherence time expected for a nuclear spin-free heterostructure. 
Lastly, we use a CPMG-64 protocol to measure the modulation of coherence due to the non-zero nuclear spin from the $^{29}\mathrm{Si}$ isotope present in the SiGe barriers surrounding the Ge quantum well. This finding points towards a non-negligible overlap of the hole wave function with $^{29}\mathrm{Si}$, possibly related Si-Ge mixing at the quantum well top Ge$\rightarrow$Si-Ge heterointerface, as observed by transmission electron microscopy in similar quantum wells~\cite{costa_reducing_2024}. 

Our study presents a framework for systematic charge and magnetic noise characterization in spin qubit devices and offers a starting point for future investigations of noise correlation to understand
the challenges of error correction schemes in noisy intermediate scale spin qubit devices. Furthermore, our findings motivate the optimization of the Ge/SiGe interface and the complete isotopic purification of the SiGe barrier layers surrounding the Ge quantum well.
Achieving nuclear spin-depleted, isotopically enriched $^{70}$Ge/$^{28}$Si$^{70}$Ge quantum wells with residual $^{73}$Ge and $^{29}$Si below 0.01\% is possible using centrifuge-separated $^{70}$GeH$_4$ and $^{28}$SiH$_4$ gas precursors~\cite{moutanabbir_nuclear_2024}. Maintaining the chemical purity of these gas precursors throughout the supply chain~\cite{sabbagh_quantum_2019} will be crucial to preserving low disorder and minimizing charge noise in future isotopically-enriched Ge spin qubit devices.
\vspace{\baselineskip}

This work was supported by the Netherlands Organisation for Scientific Research (NWO/OCW), via the Frontiers of Nanoscience program and the Open Competition Domain Science - M program.
We acknowledge support by the European Union through the IGNITE project with grant agreement No. 101069515 and the QLSI project with grant agreement No. 951852.
F.B. acknowledges support from the Dutch Research Council (NWO) via the National Growth Fund program Quantum Delta NL (Grant No. NGF.1582.22.001).
N.W.H. acknowledges support from the European Union through EIC Transition Grant
GROOVE 101113173. 
This research was sponsored in part by the Army Research Office (ARO) under Awards No. W911NF-23-1-0110. The views, conclusions, and recommendations contained in this document are those of the authors and are not necessarily endorsed nor should they be interpreted as representing the official policies, either expressed or implied, of the Army Research Office (ARO) or the U.S. Government. The U.S. Government is authorized to reproduce and distribute reprints for Government purposes notwithstanding any copyright notation herein.
This research was sponsored in part by The Netherlands Ministry of Defence under Awards No. QuBits R23/009. The views, conclusions, and recommendations contained in this document are those of the authors and are not necessarily endorsed nor should they be interpreted as representing the official policies, either expressed or implied, of The Netherlands Ministry of Defence. The Netherlands Ministry of Defence is authorized to reproduce and distribute reprints for Government purposes notwithstanding any copyright notation herein.

\section*{Author Contributions}

L.E.A.S. grew the Ge/SiGe heterostructures, developed with input from G. S., D.D.E, A.T., and D.C..
A.T.  fabricated quantum dot linear array devices, along with D.C. and contributions from D.D.E. to process development. 
S.D.O. fabricated the 2D quantum dot array device. 
L.E.A.S. performed noise measurements of the quantum dot linear array devices, with help from A.E. 
F.B., C.X.Y., B.v.S., and V.J. performed noise measurements of the 2D quantum dot array device. 
L.E.A.S. analysed the data with help from F.B. and N.W.H.. 
L.E.A.S., F.B., and G.S. wrote the manuscript with input from C.X.Y., B.v.S., N.W.H., and M.V.. 
G.S. and M.V. supervised the project, conceived by G.S.

\section*{Author Declarations}
N.W.H. is a founder, and M.V. and G.S. are founding
advisors of Groove Quantum BV, with N.W.H., M.V.,
and G.S. declaring equity interests. The remaining authors declare no competing interests.

\section*{Data availability}
The data sets supporting the findings of this study are openly available in 4TU Research Data at \url{https://doi.org/10.4121/33d97ca0-64be-496d-a440-218f376a9c4c}.

\section*{Methods}
\subsection{Heterostructure growth}
The Ge/SiGe heterostructure material is grown using reduced-pressure chemical vapour deposition in an ASMI Epsilon 2000 reactor. Starting from a Ge wafer a 2.5 $\upmu$m strain relaxed $\mathrm{Si}_{\mathrm{(1-x)}}\mathrm{Ge}_{\mathrm{x}}$ buffer is grown at a temperature of 800 \degree C with a final Ge concentration of 83\% using three grading steps ($1-x=0.07, 0.13, 0.17$). We lower the growth temperature to 500 \degree C for the growth of the final 200 nm of the SiGe buffer layer, the 16 nm Ge QW, the 55 nm SiGe barrier layer, and the sacrificial passivated Si cap layer. \\

\subsection{Device fabrication}
The four quantum dot devices are fabricated with multiple layers of Ti/Pd and platinum-germanosilicide (PtSiGe) ohmic contacts. These are defined by electron beam lithography and created by thermally diffusing Pt at a temperature of 400 \degree C. A first layer of Ti/Pd (3/17 nm) barrier gates are separated from the heterostructure by a 7 nm $\mathrm{Al}_2\mathrm{O}_3$ insulating oxide grown using atomic layer deposition. A second layer of Ti/Pd (3/37 nm) plunger gates are created and separated by 5 nm of $\mathrm{Al}_2\mathrm{O}_3$ from the first layer of barrier gates. Details for the fabrication of the `3-4-3' device can be found in Ref.~\cite{wang2024operating}.

\subsection{Flank method electrical characterization of quantum dots}
We cool down multi-hole quantum dots defined underneath the charge sensor or plunger gates of a device in a Leiden cryogenic dilution refrigerator operating at a mixing chamber base temperature of 70 mK. To measure the charge noise using the flank method, we first set a source-drain bias of 0.1 mV across the device and subsequently tune the surrounding gates until we measure a current of ~1 nA through the device. We define a multi-hole quantum dot underneath a sensor or plunger gate by fine tuning the barrier gates surrounding the gate of interest until we observe a spectrum of Coulomb peaks. We then measure the current $I_\mathrm{sd}$ on the left flank of the Coulomb peak where the slope $|\mathrm{d}I_\mathrm{sd}/\mathrm{d}V_\mathrm{sd}|$ of the Coulomb peak is the largest at a rate of 2 kHz for a duration of 100 s using a Keithley DMM6500 multimeter. To calculate the current power spectral density $S_{\mathrm{I}}$, we first split each 100 second current trace into 10 segments of 10 seconds each, and subsequently average over the current power spectral densities ($S_\mathrm{I} = 1/N \sum_{i=1}^{N} S_{\mathrm{I}}^i$) that we evaluate from each segment.
For each Coulomb peak that we analyse, we convert the current power spectral density $S_\mathrm{I}$ into charge noise power spectral density $S_\mathrm{\epsilon}$ using~\cite{paquelet_wuetz_reducing_2023}
\begin{equation}        S_{\mathrm{\epsilon}} = \frac{a^2S_{\mathrm{I}}}{|\mathrm{d}I_{\mathrm{sd}}/\mathrm{d}V_{\mathrm{S}}|^2}\,,
\end{equation}
where $a$ is the lever arm extracted from the analysis of the corresponding Coulomb diamond.
\subsection{Coulomb peak tracking}
On sensor 1 of device 2, we perform an ~18 h CPT experiment tracking the current $I_\mathrm{sd}$ through a quantum dot while continuously sweeping across the quantum dot using its plunger gate. For the 4 charge sensors of the `3-4-3' device we track the reflected signal of each charge sensor for ~2 hours when sweeping the sensor plunger gate voltage. We extract the position of the Coulomb peak by fitting each Coulomb peak to a hyperbolic secant function
\begin{equation}
    y = \frac{a}{\cosh^2{\left(b\,(x-x_\mathrm{0})\right)}} + c\,,
\end{equation}
where $y$ reflects the measured signal, $x$ the sensor's plunger gate voltage, $x_\mathrm{0}$ the position for which the Coulomb peak is maximum, and $a$, $b$, and $c$ are free fitting parameters. 
To calculate the voltage power spectral density $S_\mathrm{V}$ from the Coulomb peak fluctuations, we split them into 10 equal segments and for each segment we calculate the voltage power spectral density from a Fourier transformation. We find the charge noise power spectral density $S_{\mathrm{\epsilon}}$ (see Fig.\,\ref{fig:three}c) using the evaluated lever-arm of each sensor and using
\begin{equation}
    S_\mathrm{\epsilon}=a^2S_\mathrm{V}.
\end{equation}

\subsection{Voltage noise estimation on single hole quantum dots}
To estimate the voltage noise of the quantum dots where we know the exact hole occupancy we repeatedly load a hole into a dot by continuously sweeping the plunger gate voltage under which the quantum dot is defined. We keep track of the voltage for which a hole loads into a dot by fitting each voltage sweep to a sigmoid function given by

\begin{equation}
    y = \frac{a}{1+\exp\left(\frac{x-x_\mathrm{0}}{\tau}\right)} + b\,,
\end{equation}
where $y$ reflects the measured signal, $x_\mathrm{0}$ the voltage for which a hole loads into a dot, and $a$, $b$, and $\tau$ are free fitting parameters. 
We split the voltage fluctuations into 10 equal segments and calculate the voltage power spectral density $S_\mathrm{V}$ from a Fourier transform. The final voltage power spectral density is calculated from the average of the 10 segments. 

\subsection{Estimation of the relative out-of-plane angle between magnetic field and substrate}
We consider the average effective g-factor $g_\mathrm{eff} = 0.58$ of the 10 qubits~\cite{john_two-dimensional_2024} and assume in-plane principal g-tensor components of $g_x = - g_y = 0.04$~\cite{wang2024operating} and an out-of-plane component of $g_z = 11$~\cite{hendrickx2024sweet}. The effective g-factor can be written in term of the out of plane angle $\theta$ and the principal components
\begin{equation}
    g_\mathrm{eff} = \sqrt{ g^2_x \cos{\theta}^2 + g^2_z \sin{\theta}^2}.
\end{equation}
By inverting the equation, we estimate the most plausible misalignment angle of $\theta = 3^\circ$.

\subsection{Extraction of Rabi and Larmor frequency}
We extract the Larmor frequency $f_\mathrm{L}$ and Rabi frequency $f_\mathrm{R}$ by fitting the data with the following functions. 
For the Rabi frequency we use
\begin{equation}
    P = A  \sin(2\pi f_\mathrm{R} t + \phi)  \exp( -t^2 / \tau^2 ) + C\ ,
\end{equation}
where $P$ represents the measured up probability and $t$ the duration of the microwave burst. The Rabi frequency $f_\mathrm{R}$, decay time $\tau$, phase $\phi$, amplitude $A$, and offset $C$ are free fitting parameters.
The extraction of the Larmor frequency is done using
\begin{equation}
    P = A \frac{f_\mathrm{R}^2}{f_\mathrm{R}^2 + \Delta^2} \sin^2\left( 0.5 t \sqrt{f_\mathrm{R}^2 + \Delta^2} \right) + C \,.
\end{equation}
Here $\Delta=f-f_{\mathrm{L}}$ with $f$ the probed frequency range during measurement. $t$, $A$, $f_\mathrm{R}$, $C$, and $f_\mathrm{L}$ are free fitting parameters.

\subsection{Qubit noise model}
Low-frequency and hyperfine noise affecting the qubits hosted in the natural Ge/SiGe heterostructure is modeled by~\cite{hendrickx2024sweet}

\begin{equation} \label{eq:noise_model}
    S_{f_\mathrm{L}} = \frac{S_\mathrm{0}}{f} + S_\mathrm{0,hf}\exp\left(\frac{-(f-f_\mathrm{Ge-73})^2}{2\sigma^2_\mathrm{Ge-73}} \right)\,.
\end{equation}

Here, $S_\mathrm{0}$ represents the low-frequency noise component at 1 Hz and $S_\mathrm{0, hf}$ represents the effective strength of the hyperfine noise acting on the qubit. $f_\mathrm{Ge-73}=\gamma_\mathrm{Ge-73}B$ is the precession frequency of $^{73}\mathrm{Ge}$ determined by its gyromagnetic ratio $\gamma_\mathrm{Ge-73}=1.48$ MHz/T and the magnetic field $B$. $\sigma_\mathrm{Ge-73}$ represents the spread of the $^{73}\mathrm{Ge}$ precession frequencies. We follow the same fitting procedure as outlined in the methods of Ref.~\cite{hendrickx2024sweet} to extract $S_\mathrm{0}$, $S_\mathrm{0, hf}$, and $\sigma_\mathrm{Ge-73}$.\\

For the fitting of the data containing influence of the $^{29}\mathrm{Si}$ nuclear spin, we expand Eq.\,\ref{eq:noise_model} with a second gaussian peak  
\begin{equation} \label{eq:Si29}
\begin{split} 
    S_{f_\mathrm{L}} = \frac{S_\mathrm{0}}{f} & + S^{\mathrm{Ge-73}}_\mathrm{0,hf}\exp\left(\frac{-(f-f_\mathrm{Ge-73})^2}{2\sigma^2_\mathrm{Ge-73}} \right) \\ 
    & + S^{\mathrm{Si-29}}_\mathrm{0,hf}\exp\left(\frac{-(f-f_\mathrm{Si-29})^2}{2\sigma^2_\mathrm{Si-29}} \right)\,.
\end{split}
\end{equation}
The fitting procedure is split up into a two stage process, where we first use Eq.\,\ref{eq:noise_model} to find $S^{\mathrm{Ge-73}}_\mathrm{0,hf}$ and $\sigma_\mathrm{Ge-73}$. We fix these parameters and then use Eq.\,\ref{eq:Si29} to find $S_\mathrm{0}$, $S^{\mathrm{Si-29}}_\mathrm{0,hf}$ and $\sigma_\mathrm{Si-29}$. The precession frequencies of $f_\mathrm{Ge-73}$ and $f_\mathrm{Si-29}$ are also fixed. \\

\subsection{Estimation of the hyperfine coupling constants}
Using the average extracted parameters of $S^{\mathrm{Ge-73}}_\mathrm{0,hf} = 1.1(3)\cdot10^6 \, \mathrm{Hz^2/Hz}$ and $\sigma_\mathrm{Ge-73} = 12(1)$ kHz that describe the modeled Gaussian peak in the power spectral density at the frequency $f_\mathrm{Ge-73}$, we estimate an integrated noise of the Larmor frequency fluctuations of
\begin{equation}
 \sigma_f = \sqrt{\sqrt{2\pi} \cdot S^{\mathrm{Ge-73}}_\mathrm{0,hf} \sigma_\mathrm{Ge-73}} = 180(8) \, \mathrm{kHz}.   
\end{equation}
This sets an approximate boundary for the hyperfine-limited dephasing time of $T_2^* = (\pi \sqrt{2} \sigma_f)^{-1} = 1.25(5) \, \mathrm{\upmu \, s}$, qualitatively similar to what measured experimentally in the order of $1 - 2 \, \mathrm{\upmu \, s}$~\cite{john_two-dimensional_2024}. 

We use the same procedure to assess the influence of the interaction with the Si-29 nuclei in the barrier to the Ge hole spin qubits.
We consider the average parameters extracted from the fits in Fig.~\ref{fig:four}, $S^{\mathrm{Si-29}}_\mathrm{0,hf} = 9.0(8) \cdot 10^3 \, \mathrm{Hz^2/Hz}$ and $\sigma_\mathrm{Si-29} = 99(20)$ kHz, that leads to an integrated noise of $\sigma_f = 47(5)$ kHz and a $^{29}$Si-limited $T_2^* = 4.8(4) \, \mathrm{\upmu s}$.

\subsection{Charge noise-limited \texorpdfstring{\textit{T}}\texorpdfstring{\boldmath{$_{2}$}}\texorpdfstring{\boldmath{$^{*}$}} ~}
We evaluate the charge noise-limited $T_2^*$ using the extracted voltage noise amplitude of $\sqrt{S_\mathrm{V}} = 12(1) \, \mathrm{\upmu V / \sqrt{Hz}}$ at 1 Hz and the effective g-factor susceptibility of $\frac{\Delta g}{\Delta V} = 6.7 \cdot 10^{-4} \, \mathrm{mV^{-1}}$. The dephasing time $T_2^*$ can be approximated in a quasi-static configuration by~\cite{struck_low-frequency_2020, paquelet_wuetz_reducing_2023, wang2024operating}
\begin{equation}
    T_2^* = \frac{1}{\sqrt{2} \pi \Delta f},
\label{eq:T2*}
\end{equation}
with $\Delta f$ the amplitude of the qubit frequency fluctuations due to a voltage noise with root main square amplitude of $\Delta V_\mathrm{RMS}$
\begin{equation}
    \Delta f = \frac{\Delta g}{\Delta V} \Delta V_\mathrm{RMS}.
\end{equation}
We compute $\Delta V_\mathrm{RMS}$ over a period of time $T$ assuming a power spectral density of $S_\mathrm{V}/f$, and integrating the noise from a low- and high-frequency cutoff values, $f_L = T^{-1}$ and $f_H$, respectively
\begin{equation}
    \Delta V_\mathrm{RMS} = \sqrt{ \int_{f_L}^{f_H} \frac{S_\mathrm{V}}{f} \, df } = \sqrt{S_\mathrm{V} \ln{\frac{f_H}{f_L}}}.
\end{equation}
By considering realistic values of $f_L = 1 $ mHz and $f_H = 1 $ MHz, we determine $\Delta V_{RMS} = 54(5) \, \mathrm{\upmu V}$. 
We then distinguish two cases. For B = 117.5 mT, the resulting fluctuations leads to $\Delta f = 60(5) \, \mathrm{kHz}$ and a charge-noise limited dephasing time of $T_2^* = 3.7(3) \, \mathrm{\upmu s}$. For a low magnetic field of 10 mT, we obtain an increased value of $T_2^* = 44(4) \, \mathrm{\upmu s}$ due to the more than tenfold reduction in $\Delta f$. \\

\end{document}


\renewcommand{\thesection}{S\arabic{section}}
\renewcommand{\thetable}{S\arabic{table}}
\renewcommand{\thefigure}{S\arabic{figure}}

\title{Exploiting epitaxial strained germanium for scaling low noise spin qubits at the micron-scale}

\author{Lucas E. A. Stehouwer$^{1}$}
\author{Cécile X. Yu$^{1}$}
\author{Barnaby van Straaten$^{1}$}
\author{Alberto Tosato$^{1}$}
\author{Valentin John$^{1}$}
\author{Davide Degli Esposti$^{1}$}
\author{Asser Elsayed$^{1}$}
\author{Davide Costa$^{1}$}
\author{Stefan D. Oosterhout$^{2}$}
\author{Nico W. Hendrickx$^{1}$}
\author{Menno Veldhorst$^{1}$}
\author{Francesco Borsoi$^{1 \dag}$}
\author{Giordano Scappucci$^{1 \dag *}$}

\affiliation{$^{1}$QuTech and Kavli Institute of Nanoscience, Delft University of Technology, Lorentzweg 1, 2628 CJ Delft, The Netherlands \\ $^{2}$QuTech and Netherlands Organization for Applied Scientific Research (TNO), Stieltjesweg 1, 2628 CK Delft, The Netherlands \\$^{\dag}$ These authors contributed equally
\\ $^*$Email: g.scappuci@tudelft.nl}
\date{\today}
\pacs{}

\maketitle
\tableofcontents
\newpage

\section{Electron temperature and lever arm extraction}
\begin{figure}[htp]
    \centering
	\includegraphics[width=176mm]{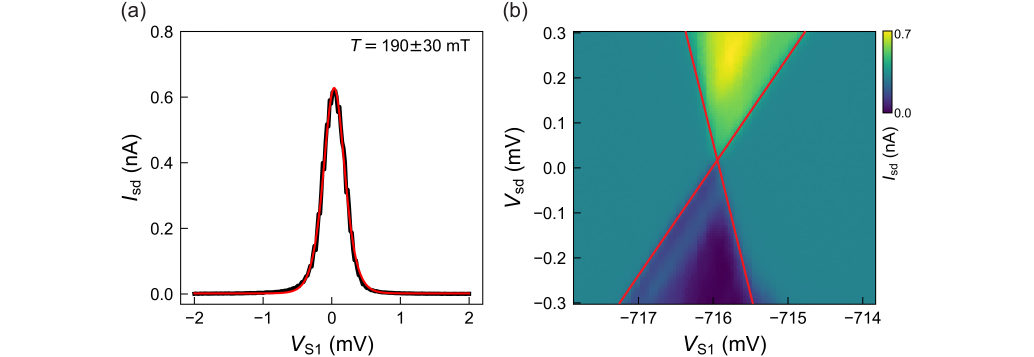}
	\caption{(a) Current $I_\mathrm{sd}$ as a function of sensor gate voltage $V_\mathrm{S1}$. We make an estimation of the electron temperature by fitting (red line) using the formula $I_\mathrm{sd} = a + b \cosh^{-2} \left( \frac{\alpha(V_\mathrm{0}-V)}{2k_\mathrm{B}T} \right)$ of the electron temperature. Here, $a$, $b$, $V_\mathrm{0}$, and $T$ are free fitting parameters, with $T$ the estimated electron temperature. $\alpha$ is the lever arm corresponding to the analysed Coulomb peak. (b) $I_\mathrm{sd}$ measured as a function of quantum dot voltage $V_\mathrm{S1}$ and source drain voltage $V_\mathrm{sd}$. We extract the lever arm from the slopes of the red lines using $\alpha = |\frac{m_\mathrm{s}m_\mathrm{d}}{m_\mathrm{s}-m_\mathrm{d}}|=0.18$ eV/V, where $m_\mathrm{s}$ and $m_\mathrm{d}$ are the slopes of the Coulomb diamond from source and drain respectively.}   
\label{fig:S1}
\end{figure}

\newpage

\section{Noise spectra from linear arrays of single hole transistors}

\begin{figure}[htp]
    \centering
	\includegraphics[width=176mm]{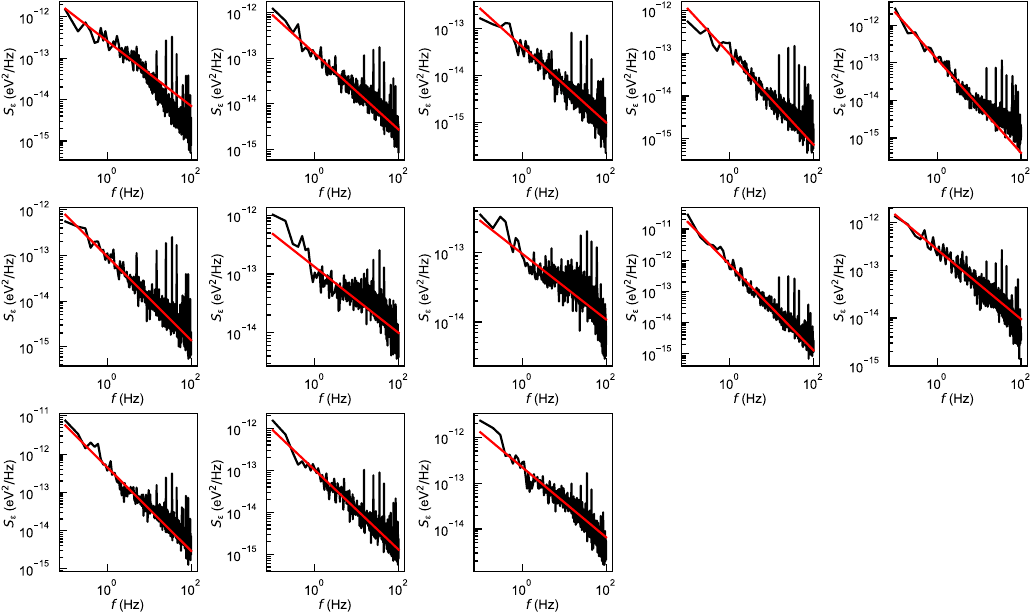}
	\caption{Ensemble of charge noise spectra $S_\mathrm{\epsilon}$ taken at different Coulomb peak flanks but unknown charge occupancy to sample the distribution of charge noise levels of quantum dot sensor 1 of device 1. Each noise spectrum is fitted to $S_0/f^{\alpha}$ (red line) using a fit range between 0.1 and 10 Hz. We extract the charge noise $\sqrt{S_0}$ at 1 Hz, which is presented in Fig.\,1e (device 1, S1) of the main text. Across the acquired noise spectra, we find an average charge noise of $S_\mathrm{0}=0.4(1)\,\mathrm{Hz}^2/\mathrm{Hz}$ and $\alpha=0.9(2)$. Measurement details can be found in the data repository (doi in main text).}  
\label{fig:S2}
\end{figure}

\newpage

\begin{figure}[htp]
    \centering
	\includegraphics[width=176mm]{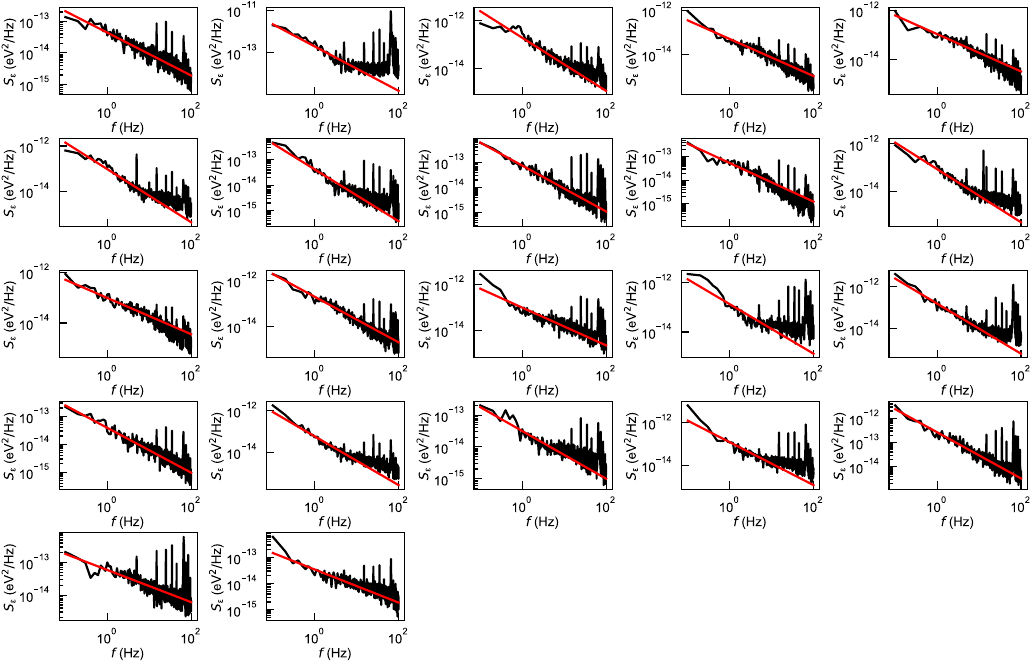}
	\caption{Ensemble of charge noise spectra $S_\mathrm{\epsilon}$ taken at different Coulomb peak flanks but unknown charge occupancy to sample the distribution of charge noise levels of quantum dot sensor 2 of device 1. Each noise spectrum is fitted to $S_0/f^{\alpha}$ (red line) using a fit range between 0.1 and 10 Hz. We extract the charge noise $\sqrt{S_0}$ at 1 Hz, which is presented in Fig.\,1e (device 1, S2) of the main text. Across the acquired noise spectra, we find an average charge noise of $S_\mathrm{0}=0.3(1)\,\mathrm{Hz}^2/\mathrm{Hz}$ and $\alpha=0.9(2)$. Measurement details can be found in the data repository (doi in main text).}  
\label{fig:S3}
\end{figure}

\newpage

\begin{figure}[htp]
    \centering
	\includegraphics[width=176mm]{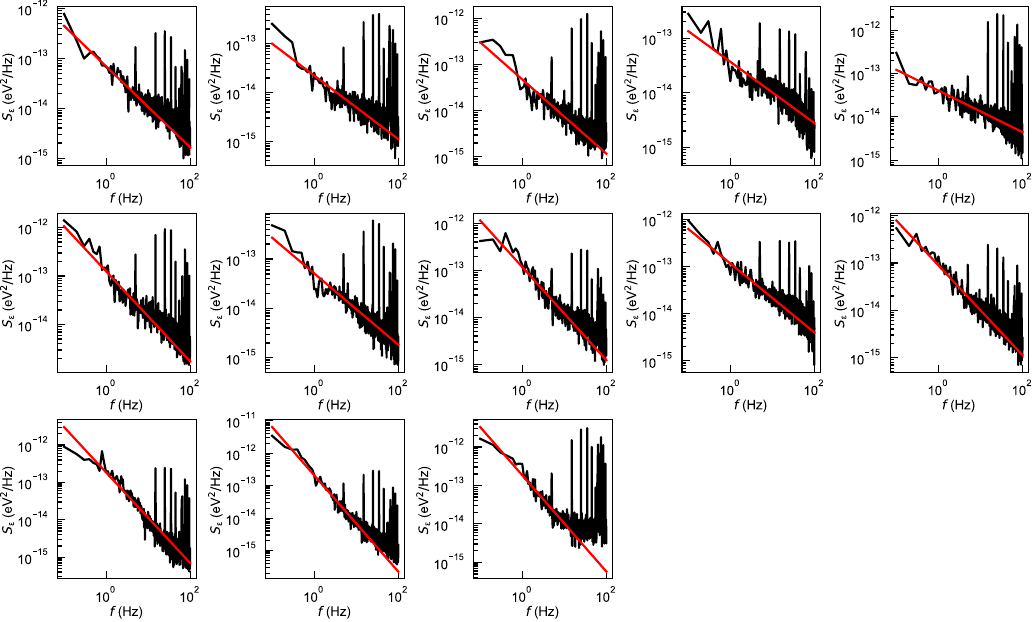}
	\caption{Ensemble of charge noise spectra $S_\mathrm{\epsilon}$ taken at different Coulomb peak flanks but unknown charge occupancy to sample the distribution of charge noise levels of quantum dot sensor 1 of device 2. Each noise spectrum is fitted to $S_0/f^{\alpha}$ (red line) using a fit range between 0.1 and 10 Hz. We extract the charge noise $\sqrt{S_0}$ at 1 Hz, which is presented in Fig.\,1e (device 2, S1) of the main text. Across the acquired noise spectra, we find an average charge noise of $S_\mathrm{0}=0.3(1)\,\mathrm{Hz}^2/\mathrm{Hz}$ and $\alpha=0.9(3)$. Measurement details can be found in the data repository (doi in main text).}  
\label{fig:S4}
\end{figure}

\newpage

\begin{figure}[htp]
    \centering
	\includegraphics[width=176mm]{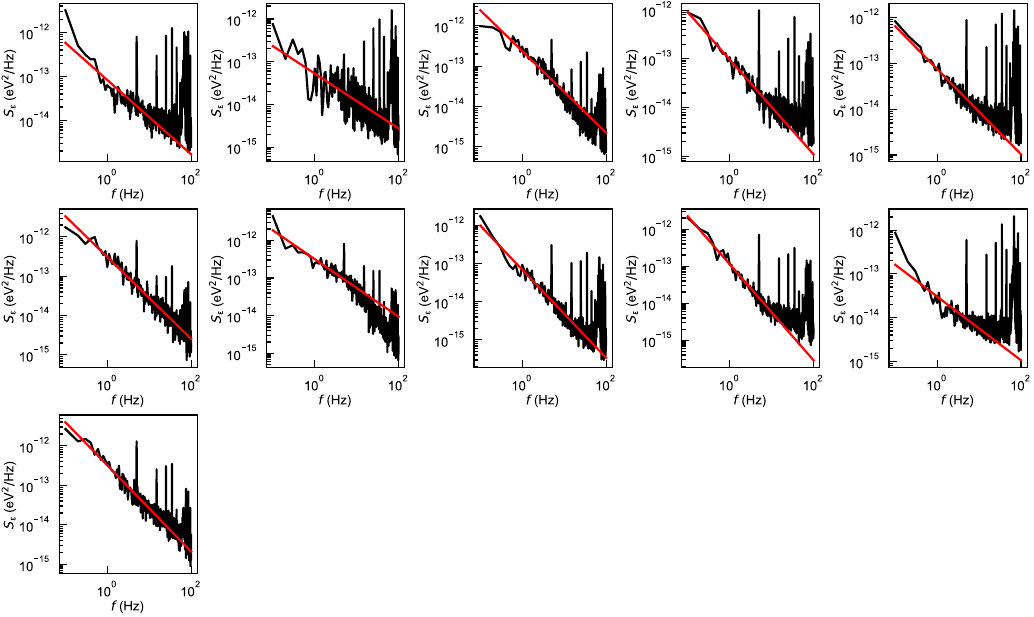}
	\caption{Ensemble of charge noise spectra $S_\mathrm{\epsilon}$ taken at different Coulomb peak flanks but unknown charge occupancy to sample the distribution of charge noise levels of quantum dot sensor 2 of device 2. Each noise spectrum is fitted to $S_0/f^{\alpha}$ (red line) using a fit range between 0.1 and 10 Hz. We extract the charge noise $\sqrt{S_0}$ at 1 Hz, which is presented in Fig.\,1e (device 2, S2) of the main text. Across the acquired noise spectra, we find an average charge noise of $S_\mathrm{0}=0.4(1)\,\mathrm{Hz}^2/\mathrm{Hz}$ and $\alpha=1.0(2)$. Measurement details can be found in the data repository (doi in main text).}  
\label{fig:S5}
\end{figure}

\newpage

\section{Charge noise from sensing dot Coulomb peak tracking}
\begin{figure}[htp]
    \centering
	\includegraphics[width=176mm]{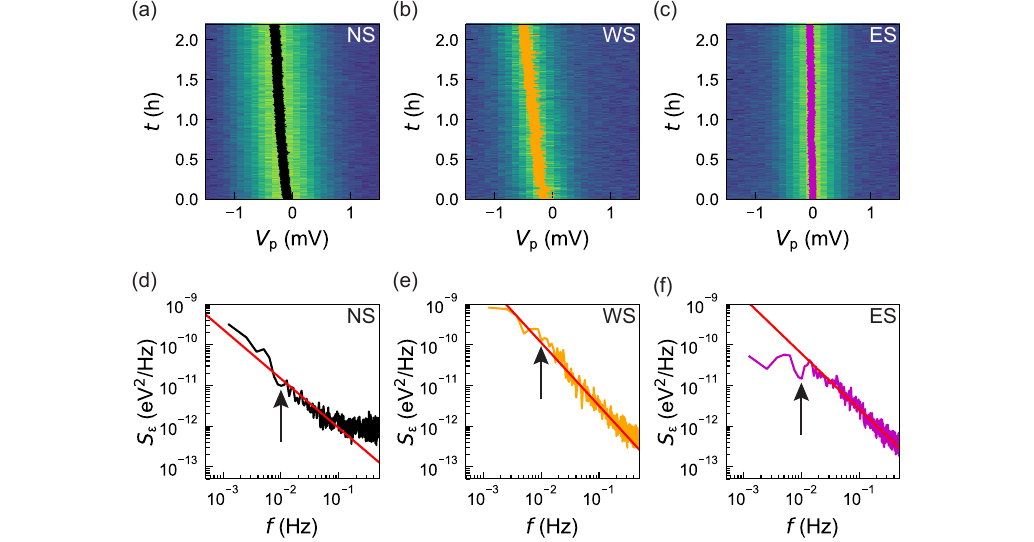}
	\caption{(a)-(c) two hour Coulomb peak tracking experiment of the north sensor (NS), west sensor (WS), and east sensor (ES) of the 3-4-3 device shown in Fig.\,2a of the main text. The black, orange, and purple lines are the estimated Coulomb peak positions. (d)-(f) calculated charge noise power spectral densities of each sensor. We extract the charge noise at a frequency of 10 mHz (black arrow) using a linear fitting procedure (red line). For Figs.\,S6(d) and (e) we use a fitting range between 1 and 100 mHz, while for Fig.\,S6(f) we use a fitting range between 10 and 150 mHz.}  
\label{fig:S6}
\end{figure}

\newpage

\section{Charge noise from hole filling experiment}
\begin{figure}[htp]
    \centering
	\includegraphics[width=176mm]{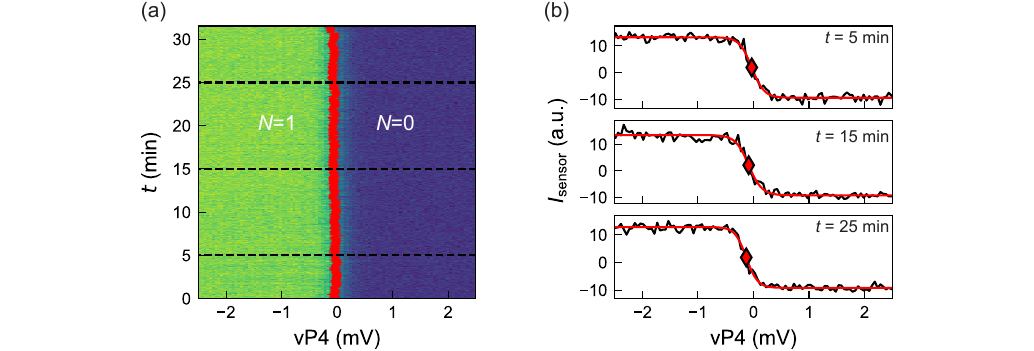}
	\caption{(a) Examplary hole filling experiment on quantum dot Q4 where we repeatedly sweep over the interdot transition filling the quantum dot with a single hole. The red line shows the estimated interdot transition. (b) Line cuts of the sweep across the transition line from (a) for $t=5$, $t=15$, and $t=25$ minutes. We estimate the position of the transition line by fitting each line cut to a sigmoid function (red) and finding the inflection point (red diamond).}  
\label{fig:S7}
\end{figure}

\newpage

\section{Spin states in a double quantum dot and parity readout}
\begin{figure}[htp]
    \centering
	\includegraphics[width=176mm]{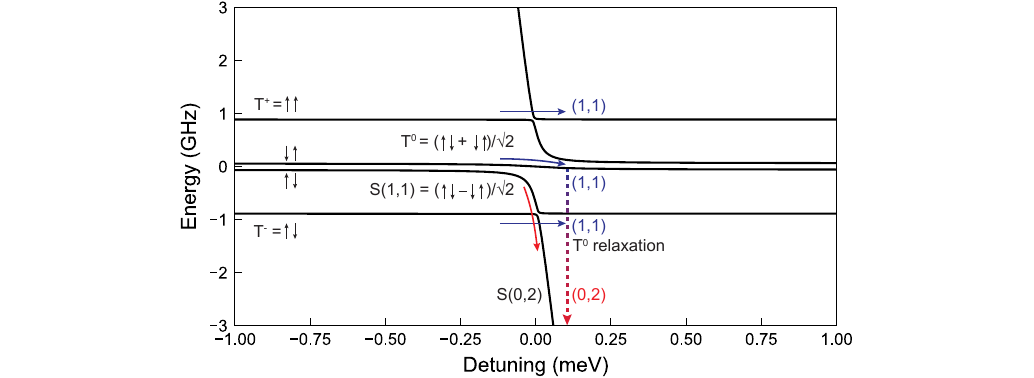}
	\caption{Energy levels of the spin states in a double quantum dot simulated assuming realistic experimental parameters: $g$-factors of 0.503 and 0.577, magnetic field of 117.5 mT, tunnel coupling $t_\mathrm{c} = 0.5 $ GHz, and a small relative quantization axis difference of $9^\circ$. The detuning energy axis is obtained by converting a typical detuning voltage window of 20 mV in electronvolt through a lever arm of 0.1 eV/V. The plot displays the readout mechanism of the four spin states. The two triplet parallel states are blocked when pulsing through the charge anticrossing (at zero detuning) and map to the (1,1) charge state. The antiparallel state $\ket{\uparrow\downarrow}$ is converted to the S(0,2) ground state. Finally, the $\ket{\downarrow\uparrow}$ spin state maps first to the $T^0$ state which relaxes to the singlet ground state S(2,0) much quicker than the typical readout time of $3-10 \, \mathrm{\upmu s}$. The basic principles of the energy anticrossings and readout mechanism in the (1,1) to (0,2) transition also apply to the (3,1) to (0,4) transition studied in the main text.}
\label{fig:S8}
\end{figure}

\newpage

\section{Charge noise from spin-echo protocols}
\begin{figure}[htp]
    \centering
	\includegraphics[width=176mm]{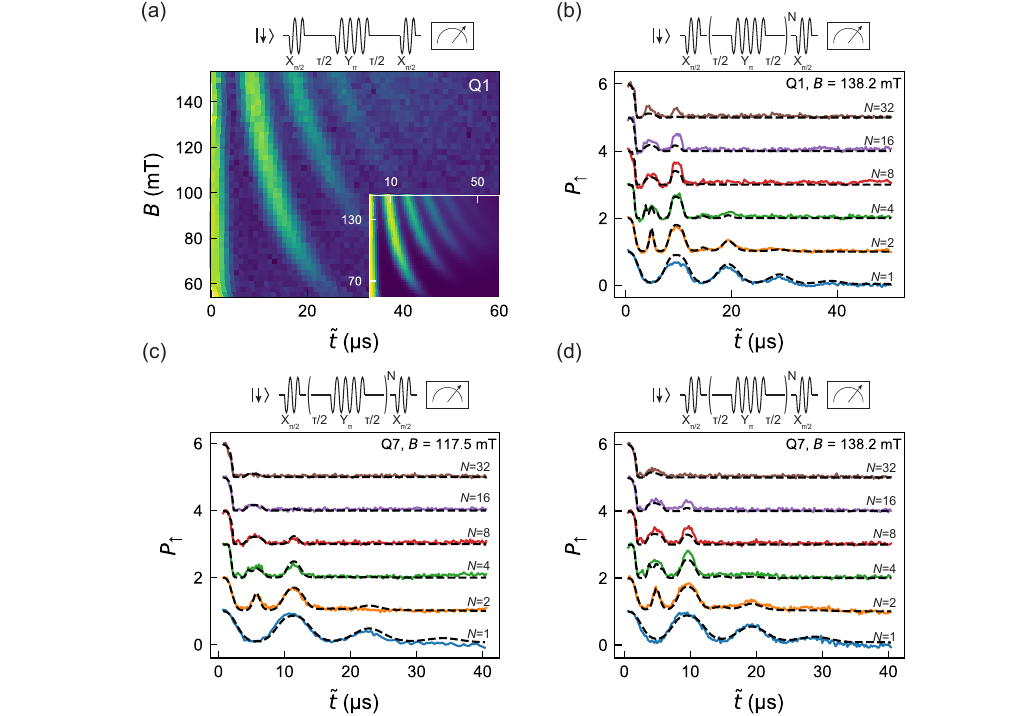}
	\caption{(a) CPMG-1 experiment performed on qubit Q1 for a range of magnetic fields $B$. The bottom right inset shows the model as presented in the Methods section G of the main text. (b) additional CPMG-$N$ data taken on qubit Q1 at a magnetic field of $B=138.2$ mT. This data, together with the data taken at $B=117.5$ mT and presented in Fig.\,3e, is used to fit to the noise model of Fig.\,3c to obtain an accurate estimation of the charge noise and hyperfine noise. 9 (c)-(d) CPMG-$N$ data taken on qubit Q7 at magnetic fields of 117.5 and 138.2 mT to estimate the charge and hyperfine noise from the fitting procedure of the noise model of Fig.\,3c as presented in the main text. In panels (b)-(d) each coherence trace is offset by unity for clarity. Dotted black lines represent the estimated coherence from the noise model.}  
\label{fig:S9}
\end{figure}

\newpage

\section{Extrapolation of Hahn dephasing time in absence of hyperfine noise}
\begin{figure}[htp]
    \centering
	\includegraphics[width=155mm]{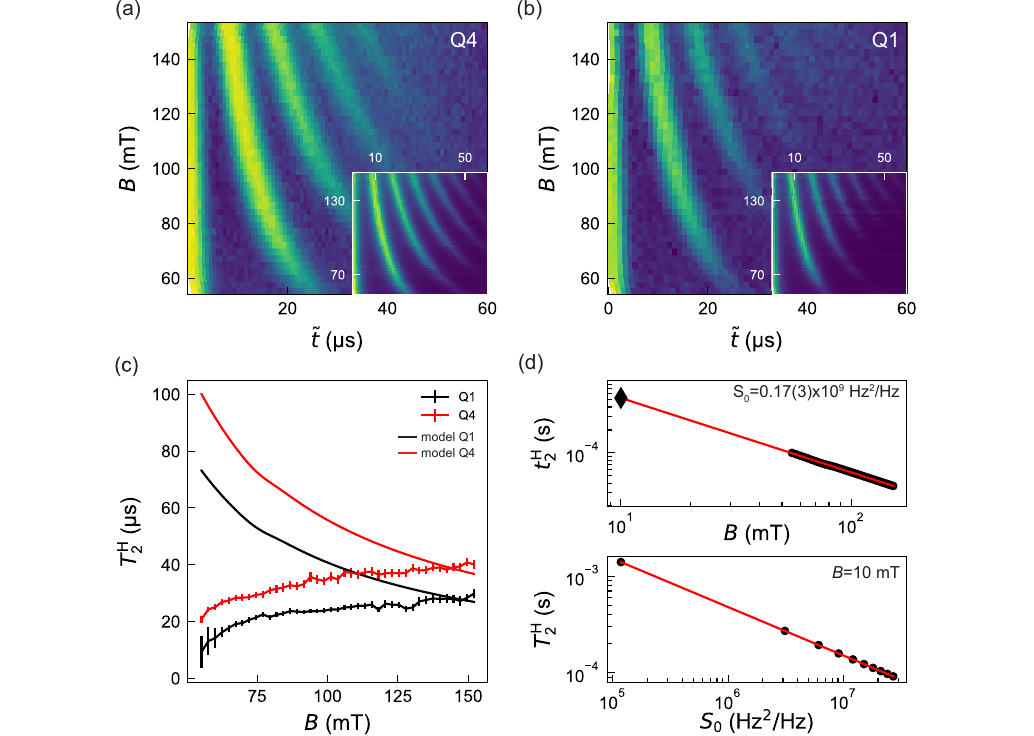}
	\caption{(a)-(b) Hahn echo coherence versus magnetic field and total wait time $\Tilde{t}$ for qubit Q4 (panel a) and qubit Q1 (panel b). For each magnetic field $B$ we extract the Hahn echo coherence time $T_\mathrm{2}^{\mathrm{H}}$ using the fitting formula $A\exp(-(\tau/t_\mathrm{2}^\mathrm{H})^\alpha) / (1 - a_\mathrm{0}\cos(\pi \gamma_{Ge-73} B \tau))^2+B$, where $A$, $B$, $\alpha$, $t_\mathrm{2}^\mathrm{H}$, and $a_\mathrm{0}$ are used as free fitting parameters. (c) We plot the extracted $t_\mathrm{2}^\mathrm{H}$ as a function of magnetic field $B$ for qubit Q1 and qubit Q4. Additionally, using the noise model from Fig.\,3 of the main text, we model the coherence for qubits Q1 and Q4 when setting $S_\mathrm{0,hf}=0$ $\mathrm{Hz}^\mathrm{2}/\mathrm{Hz}$ and extract the Hahn echo coherence time $t_\mathrm{2}^\mathrm{H}$ for this idealised case. The solid lines show the extracted coherence times for qubit Q1 (black) and qubit Q4 (red). For both qubits we see a cross over point around 150 mT below which the hyperfine noise starts to dominate over the charge noise. It exemplifies the possible increase in coherence time when isotopically purifying the material stack. (d) We take the average charge noise of $S_\mathrm{0}=0.17(3)\,\mathrm{Hz}^\mathrm{2}/\mathrm{Hz}$ from the analysed qubits Q1, Q4, and Q7. Similar to (c), we set the hyperfine noise level to 0 and extract Hahn echo coherence time $t_\mathrm{2}^\mathrm{H}$. We extrapolate $t_\mathrm{2}^\mathrm{H}$ to a low magnetic field of 10 mT and find a Hahn echo coherence time of around 0.4 ms. In the bottom panel we fix the magnetic field to 10 mT and investigate the behaviour of $t_\mathrm{2}^\mathrm{H}$ as a function of charge noise $S_\mathrm{0}$. The panels show how $t_\mathrm{2}^\mathrm{H}$ could be improved towards 1 ms when addressing the two main noise sources present in the material stack being hyperfine noise and charge noise.  
}
\label{fig:S10}
\end{figure}